\documentclass[preprint,12pt]{elsarticle}
\usepackage{amsfonts}
\usepackage{amsmath}
\usepackage{psfrag}
\usepackage{graphicx}
\usepackage{amssymb}
\usepackage{subfigure}
\usepackage{algorithmic}
\usepackage{algorithm}
\usepackage{color,soul}
\usepackage{booktabs}
\usepackage{threeparttable}%%%LBL
\usepackage{multirow}
\usepackage{lineno,hyperref}
\usepackage{ulem}
\usepackage{times}
\usepackage{lscape}

\newproof{pot}{Proof}

\soulregister\cite7 \soulregister\ref7 \journal{Journal of
Neuroscience Methods}
\begin{document}
\begin{frontmatter}

\title{Motor-imagery classification model for brain-computer
interface: a sparse group filter bank representation model}
\author{Cancheng Li$^{1,2}$\corref{}}
\author{Chuanbo Qin$^{3}$\corref{}}
\author{Jing Fang$^{4}$\corref{cor1}}
\address{$^{1}$School of Biological Science and Medical Engineering, Beihang University, Beijing, China}
\address{$^{2}$School of Information Science and Engineering, Lanzhou University, Lanzhou, China}
\address{$^{3}$Department of Intelligent Manufacturing, Wuyi University, Jiangmen, China}
\address{$^{4}$Shenzhen Institutes of Advanced Technology, Chinese Academy of Sciences, Shenzhen, China}
\cortext[cor1]{Corresponding author. \\{\it E-mail addresses:}
jing.fang@siat.ac.cn}

\begin{abstract}
{\it Background:} Common spatial pattern (CSP) has been widely used
for feature extraction in the case of motor imagery (MI)
electroencephalogram (EEG) recordings and in MI classification of
brain-computer interface (BCI) applications. BCI usually requires
relatively long EEG data for reliable classifier training. More
specifically, before using general spatial patterns for feature
extraction, a training dictionary from two different classes is used
to construct a compound dictionary matrix, and the representation of
the test samples in the filter band is estimated as a linear
combination of the columns in the dictionary matrix.

\noindent {\it New method:} To alleviate the problem of sparse small
sample (SS) between frequency bands. We propose a novel sparse group
filter bank model (SGFB) for motor imagery in BCI system.

\noindent {\it Results:} We perform a task by representing residuals
based on the categories corresponding to the non-zero correlation
coefficients. Besides, we also perform joint sparse optimization
with constrained filter bands in three different time windows to
extract robust CSP features in a multi-task learning framework. To
verify the effectiveness of our model, we conduct an experiment on
the public EEG dataset of BCI competition to compare it with other
competitive methods.

\noindent {\it Comparison with existing methods:} Decent
classification performance for different subbands confirms that our
algorithm is a promising candidate for improving MI-based BCI
performance.

\end{abstract}

\end{frontmatter}

\section{Introduction}\label{sec.introduction}
Brain-computer interface (BCI) is a new method for
the communication and control between the human brain and external
devices \cite{Chaudhary2016,Ang2015}. Compared with the evoked
potential-based BCI, motor imagery (MI) is easily operated and does
not rely on external stimuli \cite{Guger2000,Yang2017}. It has been
suggested to be suitable for the mechanical control and exercise
rehabilitation training. MI system shows various applications, such
as controlling the movement of a wheelchair, the mouse cursor on the
computer screen, and the movement of the left and right direction by
imagining the left and right hands
\cite{Bozinovski1988,Blankertz2008}. Electroencephalogram (EEG)
signal of the brain activities can effectively control the execution
of MI tasks. In the past decade, its low cost, non-invasiveness and
wide availability have attracted the interest of many researchers
\cite{Vidaurre2010}.

At present, the widely used EEG signals for BCI system control
include sensorimotor rhythms (SMRs), event-related potentials (ERP),
and visual evoked potentials (VEP)
\cite{Zhang2018,Ang2012,Park2017,Ang2012a,Gu2013}. Particularly,
event-related desynchronization / synchronization (ERD / ERS)
utilizes the $mu$ rhythm power of sensory motor rhythm (SMR).
Recently, a series of BCI has been established based on rhythmic
activity recorded on the sensorimotor cortex. SMR draws the
attention from BCI using non-invasive neural recording like EEG. SMR
is a feature as a band power change within a particular EEG
frequency band {\color{blue}\cite{Guger2000}}. At the same time, the
EEG band appears in the brain region of sensory organ motion
imaging. Therefore, the EEG power conversion can be correlated as
the control of MI task.

Until now, many algorithms have been applied to the EEG
classification in BCI system
\cite{Park2017,Higashi2013,Lotte2011,Arvaneh2011,Siuly2012,Fu2020,Zuo2020}.
For example, a common spatial pattern (CSP) as a classic feature
extraction method was introduced in this field. It uses the spatial
features of ERD/ERS, which consists of a spatial filtering technique
and simultaneously detects filters maximizing the variance for one
class and minimizing the variance for another class \cite{Li2011aa}.
The CSP is greatly effective to classify motor imagery EEG, since
the variance of the bandpass filtered signals is equal to the
bandpower. This is why CSP plays a key role in discriminating the
information of SMR related EEG data \cite{Thomas2009,Wu2015aa}.
Besides, algorithms combining filter-bank structure with
regularization common spatial pattern (RCSP) have also been studied
by many researchers. In \cite{Park2017}, Park {\it et al.} presented
RCSP feature extraction on each frequency band, and used mutual
information as the individual feature algorithm in a small sample
(SS). Moreover, a sub-band regularized common spatial pattern
(SBRCSP) used principal component analysis (PCA) to extract RCSP
features from all frequency sub-bands \cite{Ang2012}. In
\cite{Park2017}, the author used the filter-bank regularized common
spatial pattern (FBRCSP) selected optimum frequency bands for
extracting mutual information of RCSP feature. Apart from the
methods of feature extraction, another research interest in this
field is on the complex classification algorithm to improve the
accuracy of EEG classification and provide good robustness
\cite{Chen2018a,Kee2017,Peng2021,Fang2019,Wang2020,Nguyen2015,
Arvaneh2014,Hu2013aa,Meziani2019,Miao2017}. In earlier studies, with
the assumption that sample covariance matrix of different classes
are similar, linear discriminant analysis (LDA) was used as the main
algorithm to distinguish the two types of motor imagination
\cite{xupeng2011,Vidaurre2011}. In order to improve the
generalization ability of the model, numerous regularization
classifications have been applied for SMR classification. For
instance, Mahanta {\it et al.} improved the reliability of the
classifier by regularized LDA to correct inaccuracy of estimated
covariance matrices \cite{Mahanta2013}. Moreover, the sparse
representation classification (SRC) which has been successfully
applied in the image field \cite{Wright2009,Li2014} is also employed
in many studies. The SRC aims to estimate the sparse representation
of the test samples as a linear combination of the columns (ie,
training samples) in the dictionary matrix. Then, we identify the
category by minimizing the reconstruction error, the labels of the
test samples are determined by detecting which class the training
samples providing the smallest residual norm belong to
\cite{Friedman2010}. Many studies show that by introducing the SRC
scheme, the classification accuracy on SMR can be distinctly
improved \cite{Li2012,Zhang2017a,Zhang2015,Wu2015qq,Shin2012}.

Many previous researchers focused on SMR feature extraction and
pattern recognition. Then, when the dimensions of the training data
are insufficient, the number of training samples cannot truly
reflect the distribution of features, which can lead to
unsatisfactory results. In order to obtain relatively large data, a
long calibration time is required during BCI experiment acquisition,
which will affect the practicality of the system. On the other hand,
some research try to reduce the sample size as much as possible
without sacrificing classification accuracy.

In order to address the SMR classification in the SS situation and
decrease the calibration time of the BCI system, we propose a novel
sparse group filter bank representation model (SGFB). The most
compact representation of the test sample is estimated as a linear
combination of columns in the dictionary matrix. Moreover, unlike
the SRC scheme using only $L_{1}$-norm regularization, the SGFB
introduces two penalty factors ($L_{1,2}$-norm) to control the
sparsity by frequency bands, which effectively and automatically
select frequency bands from a sparse group representation of the
test samples and exclude those which provide no contribution. To be
specific, the EEG signals in the range of 4-40Hz are divided into
subbands with a range of 4Hz, i.e., 4-8Hz, 8-12Hz, 12-16Hz,
16-20Hz,$\cdots$,36-40Hz by a filter bank that consists of a
five-order butterworth bandpass filter. Furthermore, CSP is applied
to the divided signals by the filter bank. Finally, we apply the
SGFB method for the BCIs and the output of classification. The
performance of the SGFB algorithm is evaluated by the classification
accuracy for five subjects in the BCI competition III dataset IVa.
The good results we obtained suggest our method provides a new
direction for the classification of the small sample sets for MI.
Furthermore, for the filter band optimization, the EEG segmentation
time window is an important issue and has a certain interpretation
of the correlation among features \cite{Long2010}. In an MI system,
subjects is usually required to complete certain tasks. However, the
brain's response time to psychological tasks is an unknown
parameter. In the MI tasks, 0-1 s is considered to be the
preparatory stage before the mission, while the period between 3.5
and 4 seconds is a later stage.

The remainder of this paper is organized as follows. The previous
work is introduced in Section \ref{sec.related}. In Section
\ref{sec.three}, mathematical methods which include the CSP feature
extraction method, the sparse filter bank, and the filter bank group
sparse representation are introduced. In Section \ref{3.1}, a brief
description on the experimental procedures is provided. Results and
discussion are given in Section \ref{sec.ResDis}. This paper ends
with a conclusion in Section \ref{sec:Con}.

\section{Problem formulations and the proposal of SGFB model}
\label{sec.three}
\subsection{Mathematical Symbols}
To state the rest of our study more clearly, we first list the
mathematical symbols and their meanings in this paper as below. We
denote vectors and matrices as ltalian italics. Furthermore, we
define the main notations in the tab \ref{tal:1}.

\begin{table*}[!t]
\label{tal:1}
\renewcommand{\arraystretch}{1}
\centering \caption{Definitions of main notations} \label{settings1}
\newcommand{\tabincell}[2]{\begin{tabular}{@{}#1@{}}#2\end{tabular}}
\begin{tabular}{|l||l||l|}
\hline
Parameter & Description & Definition\\
\hline \hline
$\it{X} \in {\mathbb{R}^{m\times n}} $  & {\text{The EEG signal with K class}}  & $\it{{{X} = [X_1,X_2, \cdots ,X_{K}]}}$ \\
\hline
$\it{{X_K}} \in {R^{m*{n_k}}}$  &{\text{The training samples labeled}} & $\it{{{X_K} = [X_1^k,X_2^k,\cdots,X_{nk}^k]}}$ \\
&$\text{with the $k$th class}$ &\\
\hline
$\it{{W}}\in \mathbb{R}^{N\times 2M}$  &{\text{Spatial filter}}  & $\it{{W = \left[ {{w_1}, \cdots ,{w_{2M}}} \right]}}$\\
\hline
$D\in \mathbb{R}^{2M \times (N1 + N2)}$  &{\text{Composite matrix of single}}  & $D=[D_{Left}, D_{Right}]$  \\
&{\text{frequency band}}&\\
\hline
$\it{\widetilde D_{f}} \in {\mathbb{R}^{2M\times F}}$  &{\text{The composite matrix with 8}}  & $\it{\widetilde{D}_f =[D_{f1},D_{f2},\cdots,D_{f9}]}$ \\
&{\text{frequency bands}}  &   \\
\hline
$\it{y} \in {\mathbb{R}^{m \times 1}}$  & \text{Test sample}  &${y} = [y_{1},y_{2}, \cdots,y_{n}]$\\
\hline
$\it{{\widetilde D}} \in \mathbb{R}^{2M\times i}$  & \text{Component dictionary matrix}  &$\it{{\widetilde D =\left[ {{D_{l,1}},{D_{l,2}}, \cdots ,{D_{l,{N_l}}}} \right]}}$\\
\hline
$\it{u^{*}}$  & \text{Band sparse weight}  &$\it{{u^{*}}= \left[ {u_1^*,u_2^*, \cdots ,u_n^*}\right]}$\\
\hline
$\it{{X_L},{X_R}}$  & {\text{Left and right motor-}} &$\it{{X_L},{X_R}} \in {\mathbb{R}^{2n \times T}}$\\
&{\text{imagery signal}}&\\
\hline
$\it{X_L^{CSP},X_R^{CSP}} $  & \text{Left and right motor-imagery}  &$\it{X_L^{CSP},X_R^{CSP}} \in {\mathbb{R}^{2n\times T}}$\\
&\text{signal extracted by CSP} &\\
\hline
$\it{{\delta _l}}:{\mathbb{R}^N} \to {\mathbb{R}^N}$ &\text{Characteristic function}  &\text{-}\\
\hline
\end{tabular}
\end{table*}

\subsection{Common spatial pattern (CSP)}
\label{sec.CSP} We assume an EEG epoch with $N_{t}$ time samples
from $N_{n}$ channels. This EEG signal first passes through a set of
$N_{f}$ bandpass filters. Denote the data set of training samples
labeled with the $k$th class as
\begin{equation}
\label{equ:1} {\it{{X_k} = [X_1^k,X_2^k, \cdots ,X_{nk}^k]}} \in
{\mathbb{R}^{m\times{n_k}}}
\end{equation}
where each element is a training sample, $m$ is the dimension of the
feature space and $n_{k}$ is the total number of $X_{k}$. It
supposes that the dictionary define the samples to be classified as
${\it{{X} = [X_1,X_2, \cdots ,X_{K}]}} \in {\mathbb{R}^{m*n}}$, $K$
denotes the number of class and $n = \sum\nolimits_{i = 1}^K
{{n_i}}$. Given a query sample $\it{y} \in {\mathbb{R}^m}$, the task
of pattern recognition is to determine which class {\it{$y$}}
belongs to.

Matrix $\it{X_{j}}$ refers to the EEG signal filtered in the
$j^{th}$ filter bank for a single trial, the sample covariance
matrix $\it{C}$ for $\it{X_{j}}$ is calculated as
\begin{equation}
\label{equ:2} C = \it{\frac{{{X_j}X_j^T}}{{tr\left( {{X_j}X_j^T}
\right)}}}
\end{equation}
Suppose the EEG data obtained from a temporal interval in the $ith$
trial of class $l$ ($l = -1~or~ +1$) and assume that the EEG samples
have removed the mean within a given frequency band, then the
spatial covariance matrix of the category $l$ can be calculated as

\begin{equation} \label{equ:3} \it{\sum\nolimits_{l}^{}
= \frac{1}{{{N_l}}}\sum\nolimits_{i = 1}^{N{}_l} {{X_{i,l}}}
X_{i,l}^T}
\end{equation}
where $N_{l}$ is the number of trials belonging to the class $l$,
and $T$ denotes the transpose operator. CSP aims at seeking spatial
filters (i.e., linear transforms) to maximize the ratio of
transformed data variance between two classes as

\begin{equation}
\label{equ:4} \it{\mathop {\max }\limits_w J\left( w \right) =
\frac{{{w^{\rm T}}\sum\nolimits_1 w }}{{{w^{\rm T}}\sum\nolimits_2 w
}}~~~~~~~{\rm{ s}}{\rm{.t}}{\rm{.  }}~~~~{\left\| w \right\|_2}} = 1
\end{equation}
where ${\it{w}\in \mathbb{R}^{C}}$ is a spatial filter and
$\|{\cdot}\|$ is the $l_{2}$-norm. The CSP filter matrix {\it{$W$}}
consists of the column vector $\it{{w_i}}$. This can be achieved by
equivalently solving a generalized eigenvalue problem
$\sum\nolimits_1 {\it{w}}  = \lambda \sum\nolimits_2 {\it{w}}$. By
collecting eigenvectors corresponding to the largest and smallest
generalized eigenvalues of the $M^{*}$, a set of spatial filters are
obtained from $\it{W}$. For a given EEG sample ${\it{X}}$, the
feature vector is formed as ${\it{Z}}$ with entries as follow,

\begin{equation}
\label{equ:5} {{\rm Z}_m} = \log \left( {{\mathop{\rm var}} \left(
{w_m^TX} \right)} \right)~~~~~~~{\rm{     m = 1,}} \cdots
{\rm{,2M^{*}}}
\end{equation}
where var($\cdot$) denotes the variance.

Given two types of EEG training signals, ${\it{X_L}}$ and
${\it{X_R}}$, we define the CSP filtered signal as
\begin{equation}
\label{equ:6}
\begin{array}{l}
{\it{X_R^{CSP}}}  = W_{CSP}^T{X_R}\\
{\it{X_L^{CSP}}}  = W_{CSP}^T{X_L}
\end{array}
\end{equation}

The EEG samples ${\it{{X_{i,l}}}}$ is the bandpass filtered on
different bands instead of just a single frequency band, the CSP
features can be extracted on each frequency band. Hence, the
dimensionality of the feature vector is expanded to be 2M$\times$ N
by concatenating all of the features. Note that, the successful
application of CSP highly depends on the selection of filter bands.
The optimal filter band is generally subject-specific. An increasing
number of studies have suggested that the accuracy of MI
classification can be significantly improved by the optimization of
the filter band.

\begin{figure*}
\centering
\includegraphics[scale=0.185]{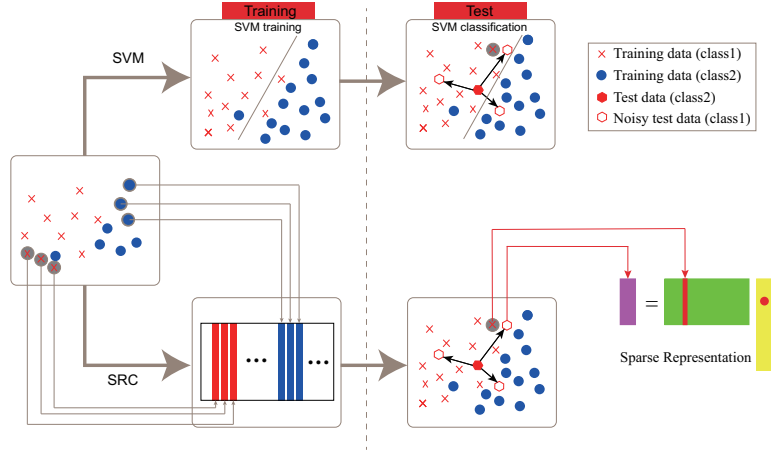}
\caption{The block diagram of the SVM and sparse representation
method.} \label{fig:1}
\end{figure*}

\subsection{Sparse representation classification (SRC)} \label{sec.related}
We have mentioned that $X$ is the training sample. In addition,
$\it{X_{ij}}$ and ${n_i}$ represent the $j$th training sample and
the number of training samples from the $i$th class respectively.
Furthermore, the test signal $\it{y} \in \mathbb{R}^{m}$ can be
sparsely represented as a linear combination of some columns of
$\it{X}$, where ${\vartheta} \in {\mathbb{R}^n}$ is a coefficient
vector corresponding to $\it{X}$. For a test sample ${\vartheta} \in
{\mathbb{R}^n}$, its representation under all the training samples
is
\begin{equation}
\label{equ:7}
\begin{array}{l}
y = {X_1}{\vartheta _1} + {X_2}{\vartheta _2} +  \cdots  + {X_c}{\vartheta _c}\\
{\rm{  }} = {X_{11}}{\vartheta _{11}} + {X_{12}}{\vartheta _{12}} +  \cdots  + {X_{c{n_c}}}{\vartheta _{c{n_c}}}\\
{\rm{  }} = X\vartheta
\end{array}
\end{equation}
Therefore, the sparse solution of Eq. \ref{equ:7} can be represented
by Eq. \ref{equ:8}.

\begin{equation} \label{equ:8}
\mathop {\arg \min }\limits_{\vartheta} \left\| {\varphi  -
Z\vartheta } \right\|_2^2 + \lambda f\left( {{{\left\| \vartheta
\right\|}_p}} \right)
\end{equation}
where $\varphi$ and $Z$ are the test sample and the training
samples, respectively. $f\left( {{{\left\| \vartheta \right\|}_p}}
\right)$ serves as an adjustable penalty function of $l_{p}$-norm
constrained, and $\lambda> 0$ is a limiting parameter, a larger
$\lambda$ can obtain a greater degree of sparse solution. According
to the changes in  $\it{Z}$, $\it{\varphi}$ and $p$, when Equation
\ref{equ:9} is minimized, we get the sparse ground coefficient
vector. $\it{{\vartheta ^*}}$. Let $\it{{\delta _i}\left(
{{\vartheta ^*}} \right)}$ be a vector whose only nonzero entries
are associated with class $i$, the class label of $\it{y}$ can be
decided as $\it{y}$ that gives the minimum reconstruction error,
i.e.,

\begin{equation}
\label{equ:9} y =
\mathop {\arg \min }\limits_{\vartheta} {\left\| {\varphi - Z{\delta
_i}\left( {{\vartheta ^*}} \right)} \right\|_2}
\end{equation}

Besides, we show the commonly used SVM and sparse representation
frameworks using the BCI in the Fig. \ref{fig:1}.

\begin{figure*}
\includegraphics[scale=0.082]{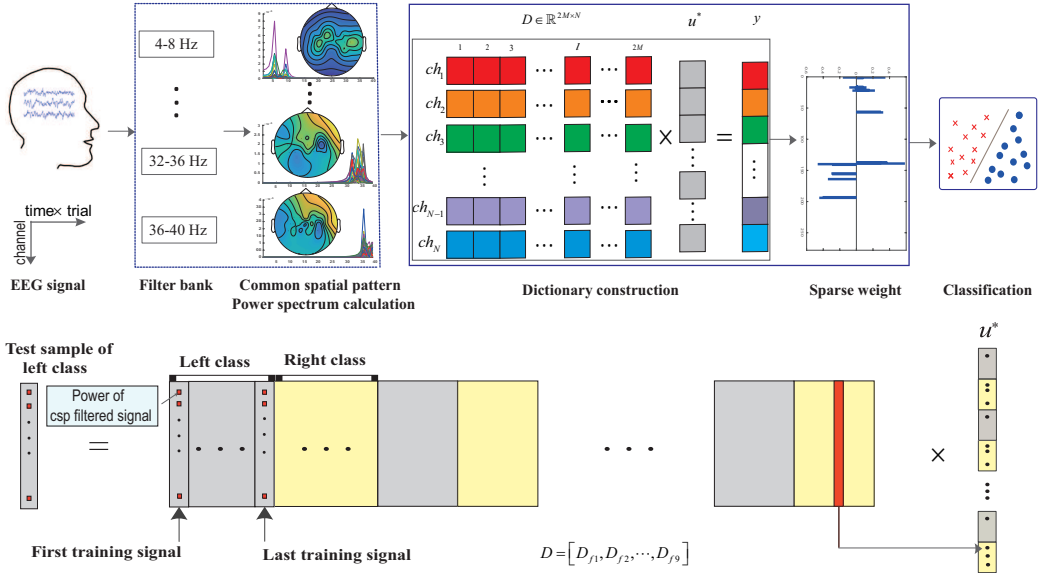}
\centering \caption{The block diagram of the filter bank group
sparse representation method. We define a component dictionary
matrix $\widetilde D = \left[ {{D_{l,1}},{D_{l,2}}, \cdots
,{D_{l,{N_l}}}} \right]$, and the $\widetilde{D}$ is a matirx of
size $2M \times N$, $u^{*}$ is of size $N \times 1$ and is divided
into $K+1$ non-overlapping group ${u_0},{u_1},{u_2}, \cdots
,{u_K}$.} \label{fig:2}
\end{figure*}

\subsection{Sparse group filter bank (SGFB)}
\label{sec.FBGSSP} Our proposed method is based on the SGFB
framework, i.e., we use the different frequency bands (4-8Hz,
8-12Hz, ... , 36-40Hz). The block diagram of the filter bank group
sparse representation method is shown in Fig. \ref{fig:2}. SRC is
used in many applications \cite{peng2019}, especially in face
recognition. SGFB is an extended approach of the SRC, which can
provide the new idea in MI classification tasks. The CSP feature of
the test samples are in the form of a linear combination and are
superior to the LDA method in the MI classification of BCI. In our
algorithm, we design the dictionary matrix to represent the test
samples.

Although many algorithms have achieved significant results in MI
classification, it is difficult to provide satisfactory results when
the size of available train samples is small. Furthermore, to
achieve better classification performance, relatively large system
calibration time is required for larger EEG data. Therefore, to
avoid sacrificing classification accuracy and obtain significant
results with small data, we propose a new classification framework
based on SGFB.

Suppose that the number of training samples for each class $l$ is
$N_{l}$ ($l$ = 1 for the left hand, and the $l = 2$ for the right
hand). Denote {\it{$\widetilde D = \left[ {{D_{l,1}},{D_{l,2}},
\cdots ,{D_{l,{N_l}}}} \right]$}} as the combined dictionary matrix
of class $l$, where the column vectors $\it{{D_{l,i}}} \in
{\mathbb{R}^{2M \times 1}}$ $(i=1,2,\cdots,N_{l})$ are the CSP
features obtained by \ref{equ:4}. Then, with the constructed
dictionary matrix, we will use SGFB to find the best sparse
representation vector {\it{$u^{*}$}}.

The coherence measures the correlation between the two component
dictionaries defined in the following way:
\begin{equation}
\label{equ:10}
M\left( {{D_L},{D_R}} \right) \buildrel \Delta \over = \max \left\{
{\left| {\left\langle {{D_{L,j}},{D_{R,k}}} \right\rangle }
\right|:j,k = 1,2, \cdots ,{N_l}} \right\}
\end{equation}

The vector ${D_{L,j}}$ is the $j^{th}$ column of ${D_L}$; similarly,
${D_{R,k}}$ is the $k^{th}$ column of ${D_R}$. The notation
$\left\langle {{D_{L,j}},{D_{R,k}}} \right\rangle$ denotes the inner
product of two vectors. We call $M$ the measure of mutual coherence
of the two component dictionaries; when $M$ is small, we say that
the complete dictionary is incoherent.

By this method, we construct a complex dictionary matrix from
different frequency bands, which includes not only the training
samples collected on the frequency band (8-13Hz) that many
researchers have already investigated, but also on the frequency
bands that are rarely considered before. As a result, good
classification effect can be obtained even for a relatively limited
amount of the training samples. Applying sparsity to the entire
training set, the SGFB estimates the optimal representation vector,
and the $\ell_{1}$ norm regularization is employed to further
exclude those insignificant frequency bands. A detailed description
of the SGFB method is given below.

We collect EEG training data from 9 different frequency bands. We
then use the CSP method for feature extraction and derive K+1
dictionaries {\it{$D_{f1}, D_{f2}, ... D_{f9}$}} in different
frequency bands, which are concatenated later into a conforming
matrix {\it{$\widetilde{D}$}}. The compounding different frequency
bands to form a composite matrix and our SGFB method is designed to
estimate the best representation vector {\it{$u^*$}} as:

\begin{equation}
\label{equ:11} {\it{{u^*}}} = \mathop {\arg \min }\limits_u
\frac{1}{2}\left\| {y - \widetilde D{u}} \right\|_F^2 + \lambda
{\left\| u \right\|_1}
\end{equation}
where $\lambda$ denotes the Hyperparameter. In order to generate a
sparse vector for the samples of $u^{*}$, the quadratic constraint
is

\begin{equation}
\label{equ:12} \mathop {\arg \min }\limits_{\it{u}} \sum\limits_{i =
1}^n {\left\| {{\it{{u_i}}} - \frac{1}{n}\sum\limits_{j = 1}^n
{{\it{{u_j}}}}} \right\|}_2^2
\end{equation}

To make the sparse vector as close as possible to its concentrated
vector, we update Eq. \ref{equ:11} as shown in Eq. \ref{equ:13}.

\begin{equation}
\label{equ:13} \mathop {\arg \min }\limits_u \frac{1}{2}\left\| {y -
\widetilde D{u}} \right\|_F^2 + \lambda {\left\| u \right\|_1} +
\frac{{{\lambda _1}}}{2}\sum\limits_{i = 1}^n {\left\| {{u_i} -
\frac{1}{n}\sum\limits_{j = 1}^n {{u_j}} } \right\|_2^2}
\end{equation}

The hyperparameter $\lambda_{1}$ is the constraint parameter, which
improves the generalization ability of the model by controlling the
sparsity of {\it{$u^{*}$}} and the sparsity between groups of these
subbands. The optimal sparse matrix {\it{${u^*}$}} can generate not
only similar sparse decompositions, but also similar weights.
Therefore, if {\it{$u_{ik}^*$}} and {\it{$u_{jk}^*$}} are not zero,
they should be with the same sign, namely, {\it{$u_{ik}^*\times
u_{jk}^*> 0$}}.

Given data {\it $y$}, {\it{$\widetilde D_{f}$}}, and the parameters
$\lambda$ and $\lambda_{1}$, if {\it{$u_{ik}^* \times u_{jk}^*>
0$}}, then {\it{$u_{ik}^*$}} and {\it{$u_{jk}^*$}} have
{\it{${\mathop{\rm sgn}} \left( {u_{ik}^*} \right) = {\mathop{\rm
sgn}} \left( {u_{jk}^*} \right)$}}, where ${\mathop{\rm sgn}} \left(
\bullet \right)$ denotes the sign function, Eq. \ref{equ:11} is
transformed into
\begin{equation}
\label{equ:14}  L\left( {{c_i},\lambda ,{\lambda _1}} \right) =
\frac{{\rm{1}}}{{\rm{2}}}\left\| {y - \widetilde Du} \right\|_F^2 +
\lambda {\left\| {{u_i}} \right\|_1} + \frac{{{\lambda
_1}}}{{\rm{2}}}\sum\limits_{i = 1}^n {\left\| {{u_i} -
\frac{1}{n}\sum\limits_{j = 1}^n {{u_j}} } \right\|} _2^2
\end{equation}

in this case, {\it{$u_i^*$}} should satisfy $(if\left(
{\it{{u_{{i_k}}}}^* \ne 0} \right))$

\begin{equation}
\label{equ:15}\frac{{\partial L\left( {{u_i},\lambda ,{\lambda _1}}
\right)}}{{\partial {u_{ik}}}}\left| \begin{array}{l}
\\
{u_i} = u_i^*
\end{array} \right. = 0
\end{equation}

It obtains
\begin{equation}
\label{equ:16}  \begin{array}{l}
 - \alpha _k^T\left( {{y_i} - \widetilde Du_i^*} \right) + \lambda {\mathop{\rm sgn}} \left( {u_{ik}^*} \right) + \\
{\lambda _1}\left( {1 - \frac{1}{n}} \right)\left( {u_{ik}^* -
\frac{1}{n}\sum\nolimits_{t = 1}^n {u_{tk}^*} } \right) = 0
\end{array}
\end{equation}

\begin{equation}
\label{equ:17} \begin{array}{l}
 - \alpha _k^T\left( {{y_j} - \widetilde Du_j^*} \right) + \lambda {\mathop{\rm sgn}} \left( {u_{jk}^*} \right) + \\
{\lambda _1}\left( {1 - \frac{1}{n}} \right)\left( {u_{jk}^* -
\frac{1}{n}\sum\nolimits_{t = 1}^n {u_{tk}^*} } \right) = 0
\end{array}
\end{equation}

Subtract equation \ref{equ:16} and \ref{equ:17} as follow,
\begin{equation}
\label{equ:18} \alpha _k^T\left( {{r_j} - {r_i}} \right) + {\lambda
_1}\left( {1 - \frac{1}{n}} \right)\left( {u_{ik}^* - u_{jk}^*}
\right) = 0
\end{equation}

which gives an equivalent form as
\begin{equation}
\label{equ:19}u_{ik}^* - u_{jk}^* = \frac{n}{{\left( {n - 1}
\right){\lambda _1}}}\alpha _k^T\left( {{r_i} - {r_j}} \right)
\end{equation}
where ${y_i} - \widetilde Du_i^*$ is the residual vector. We can
update the Eq. \ref{equ:19} and implies that

\begin{equation}
\label{equ:20}
\begin{array}{l}
\left| {u_{{i_k}}^* - u_{{j_k}}^*} \right| = \frac{n}{{\left( {n - 1} \right){\lambda _1}}}\left| {\alpha _k^T\left( {{r_i} - {r_j}} \right)} \right|\\
~~~~~~~~~~~~~~\le \frac{n}{{\left( {n - 1} \right){\lambda _1}}}{\left\| {{\alpha _k}} \right\|_2}{\left\| {{r_i} - {r_j}} \right\|_2}\\
~~~~~~~~~~~~~~\le \frac{1}{{\left( {n - 1} \right){\lambda
_1}}}{\left\| {{r_i} - {r_j}} \right\|_2}
\end{array}
\end{equation}

For calculating the optimal solution, we have rewritten Eq.
\ref{equ:12} as Eq. \ref{equ:21}
\begin{equation}
\label{equ:21}\begin{array}{l} {\rm{   }}\sum\limits_{i = 1}^n
{\left\| {{u_i} - \frac{1}{n}\sum\limits_{j = 1}^n {{u_j}} }
\right\|_2^2}\\
= \left\| {u - u\frac{1}{n}{{11}^\text{T}}} \right\|_F^2\\
 = \left\| {u\left( {I - \frac{1}{n}{{11}^T}} \right)} \right\|_F^2\\
 = Tr\left( {u{{\left( {I - \frac{1}{n}{{11}^T}} \right)}^2}{u^T}} \right)\\
 = Tr\left( {uM{u^T}} \right)
\end{array}
\end{equation}
where $Tr\left( \cdot \right)$ denotes the trace and ${\left(
\cdot\right)^T}$ denotes the transpose of a matrix, the $M = {\left(
{I - \frac{1}{T}{{11}^T}} \right)^2} \in {\mathbb{R}^{T \times
T}}$~$(I \in \mathbb{R}^{T \times T} \text{is the unit matrix and}~
1=[1,\cdots,1]^T\in \mathbb{R}^T)$. Hence, we can update the Eq.
\ref{equ:13} as
\begin{equation}
\label{equ:22} \mathop {\arg \min }\limits_i \frac{1}{2}\left\| {y -
\widetilde D{u^*}} \right\|_F^2 + \lambda {\left\| u \right\|_1} +
\frac{{{\lambda _1}}}{2}Tr\left( {uM{u^T}} \right)
\end{equation}

The $c_i$ is updated by fixing the vector ${\left\{ {{u_j}}
\right\}_{j \ne i}}$, and the optimization process is as

\begin{equation}
\label{equ:23} \mathop {\arg \min }\limits_i \frac{1}{2}\left\| {y -
\widetilde D{u_i}^*} \right\|_2^2 + \frac{{{\lambda
_1}}}{2}{M_{ii}}u_i^T{u_i} + u_i^T{h_i} + \lambda {\left\| {{u_i}}
\right\|_1}
\end{equation}
where ${h_i} = {\lambda _1}\left( {\sum\limits_{j \ne i}
{{M_{ij}}{c_j}} } \right)$. Here, we need to look for the signs of
the coefficients $c_i^j$ of $c_i$. Once all the signs of $c_i^j$ are
determined, the Eq. \ref{equ:23} can be transformed into an
unconstrained optimization problem. In this paper, we define
$h\left( {{u_i}} \right) = \frac{1}{2}\left\| {y - \widetilde
D{u_i}} \right\|_2^2 + \frac{{{\lambda _1}}}{2}{M_{ii}}u_i^T{u_i} +
u_i^T{h_i} $ and $\nabla _i^h\left| {h\left( {{u_i}} \right)}
\right| = \frac{{\partial h\left( {{u_i}} \right)}}{{\partial
u_i^j}}$

If $\widetilde D$ is large, the values in the calculated system
vector will be small. In order to solve this problem, we normalize
each column of $\widetilde D$ so that the $\ell2$ norm value of each
column is less than or equal to 1. The normalized matrix set
$\widetilde D$ is obtained by the Eq. \ref{equ:24}.

\begin{equation}
\label{equ:24} \widetilde D = \left\{ {\widetilde D \in {R^{m \times
k}}~~s.t. ~\forall j = 1, \cdots ,k,d_j^T{d_j} \le 1} \right\}
\end{equation}

In principle, we calculate the minimum reconstruction error to
determine the classification label that the test sample belongs to.
Specifically, ${\delta _l}:{\mathbb{R}^N} \to {\mathbb{R}^N}$ as a
characteristic function to select the coefficient related to class
$l$. $u^{*}$ with all non-zero coefficients are associated with the
columns of a single category $l$ in the dictionary matrix
$\widetilde{D}$. Then, the calculation of the sparse coefficient
vector is realized by the Lars algorithm\cite{lassospare}.

\begin{equation}
\label{equ:25}
\begin{array}{l}
{r_i}\left( {\overline y } \right) = {\left\| {\overline y  - {{\overline y }_l}} \right\|_2}\\
~~~~~~~{\rm{          = }}~{\left\| {\overline y  - \widetilde
D{\delta _l}\left( {{u^*}} \right)} \right\|_2}
\end{array}
\end{equation}
we approximate the test sample as ${{\it{{{\overline y
}_l}}}}={\widetilde D{\delta _l}\left( {{u^*}} \right)}$ and
calculate the residuals between ${\overline y }$ and ${{{\overline y
}_l}}$ to determine the category which the minimum reconstruction
error belongs to.

\begin{equation}
\label{equ:26}
 {\rm{class}}\left( \bar{y} \right) = \mathop {\arg \min
}\limits_i {r_i}\left({\it{\bar{y}}} \right)
\end{equation}

The column vectors in $D$ corresponding to those zero entries in
$u^{*}$ are excluded to form a optimized feature set $\tilde{D}$
that is of lower dimensionality. The given $\lambda$ and $\lambda_1$
determine the sparsity degree of $u^{*}$, correspondingly the
selection of CSP features.

For the new test sample, the corresponding subband feature vector
${D_{l,{N_l}}} \in {R^{2M \times N}}$ is estimated by CSP. According
to the sparse vector $u^{*}$, the optimal feature is selected, and
then MI classification is determined by the minimum redundancy
error.

Besides, we summarize the process of the algorithm of the SGFB
optimization and the classification process in Algorithm $1$ and
Algorithm $3$ respectively.

\section{Experimental datasets and evaluation} \label{3.1}
\subsection{EEG data description}
\label{sec.DataDes} In this paper, we use the BCI competition III
dataset IVa that is publicly available to evaluate the performance
of the proposed algorithm. More details can be found at
http://www.bbci.de/competition. These datasets are effective to
evaluate the performance of the algorithms, which has been used in
many studies.

The BCI competition III dataset IVa is available in 1,000Hz version
and 100Hz version. In this experiment, we use the 100Hz version and
1-2s interval of 3.5s motor imagery EEG. The EEG signal is obtained
from five healthy subjects (aa, al, av, aw and ay). The subjects sit
in a comfortable chair and perform motor imagery experiments. The
EEG signal is recorded by using 118 channels and 140 trials for each
class. The classes consist of the right hand and foot, i.e., the EEG
signal is provided with a total of 280 trials for each subject.

Instead of performing band-pass filtering on the original EEG
segment, we perform a set of sub-band filtering, selecting 9
sub-bands from the frequency range of 4-40Hz, and the bandpass is
4Hz, i.e., 4-8Hz, 6-10Hz, ... , 36-40Hz. Subsequently, the CSP
features can be extracted from the EEG segments of each subband.

\subsection{Experimental evaluation}
\label{sec:ev} Our experiment consists of five steps. Firstly, the
EEG signal is divided by the frequency range of 4-40Hz into 9 bands.
This filter bank consists of a bandpass filter. The range of each
band is 4-8Hz, 8-12Hz, $\cdots$,~36-40Hz respectively. Shin {\it {et
al.}} explored the effectiveness of different CSP numbers for MI
classification tasks \cite{Shin2012}. Secondly, we apply CSP to the
signal derived from the filter bank and the 32 spatial filters. As a
next, the band power is calculated in different filter banks.
Thirdly, the dictionary learning with $\ell1$ minimization is
employed in different filter bank. Finally, the label for each data
is assigned by using the ensemble.

The Classification performance is evaluated based on classification
accuracy (ACC), sensitivity (SEN), and specificity (SPE). These
statistical measures are defined as follow:

\begin{equation}
\label{equ:30} ACC=\frac{TP+TN}{TP+FP+TN+FN}
\end{equation}

\begin{equation}
\label{equ:31} SEN=\frac{TP}{TP+FN}
\end{equation}

\begin{equation}
\label{equ:32} SPE=\frac{TN}{TN+FP}
\end{equation}
where TP, TN, FP and FN denote the true positive, true negative,
false positive and false negative, respectively. Thus, ACC measures
the proportion of subjects correctly classified among all subject,
SEN and SPE correspond to the proportions of left hand and right
hand classified, respectively.

\section{Results and discussion}
\label{sec.ResDis} In the past few years, researchers have developed
a series of algorithms in MI. In this paper, We proposed a complex
model of SGFB to improve MI classification accuracy. The relevant
results are described in Table \ref{tab:performance_comparison}.
Besides, we performed a statistical comparison of the performance of
(C3-C4), CSP, FBCSP, and SGFB. Moreover, we used the 10-fold
cross-validation. We marked the average accuracy in the SGFB method
in bold for each sample. Besides, We will discuss the impact of
different parameters on classification accuracy in Section
\ref{sec:parameter}.

\begin{table*}[tp]
  \centering
  \fontsize{9.1}{9.1}\selectfont
  \caption{Procedures required by each of the compared methods.}
  \label{tab:performance_comparison}
\setlength{\tabcolsep}{0.65 mm}{
\begin{tabular}{ll}\toprule
Method      & Procedures
\\\midrule
(C3-C4)+SVM & Bandpass filtering + Variance calculation + SVM \\
CSP+SVM     & Bandpass filtering + CSP + SVM with parameter selection \\
FBCSP+SVM   & Multi-subband filtering + CSP + SVM with parameter selection\\
FBCSP+SRC   & Multi-subband filtering + SRC\\
SGFB        & Multi-subband filtering + CSP + Sparse
learning with parameter selection\\\bottomrule
\end{tabular}}
\end{table*}

To comprehensively investigate the effectiveness of our proposed
method on training data reduction, we further tested the
classification accuracy by using different numbers of training
samples from the target subject. In particular, we randomly select
30\%, 50\%, 70\% and 100\% samples (TS) from each target subject and
evaluate the classification accuracy based on repeating the above
steps 10 times. The tab. \ref{tab:4}, \ref{tab:5}, \ref{tab:6} and
\ref{tab:7} describe the average MI classification accuracy obtained
by SGFB for 30\%, 50\% 70\%, and 100\% samples, respectively. The
results show that even if only 30\% of the samples can be used for
training to obtain 81.16\% accuracy, this confirms the potential
prospect of our model in MI classification with only a few training
samples.

\begin{figure*}[t]\centering
\psfrag{t}[c][c][0.7]{$t$ (s)}
\psfrag{hc1}[c][c][0.6]{~~$\hat{c}_1$}
\psfrag{hc2}[c][c][0.6]{$\hat{c}_2$}
\subfigure[]{\includegraphics[scale=0.13]{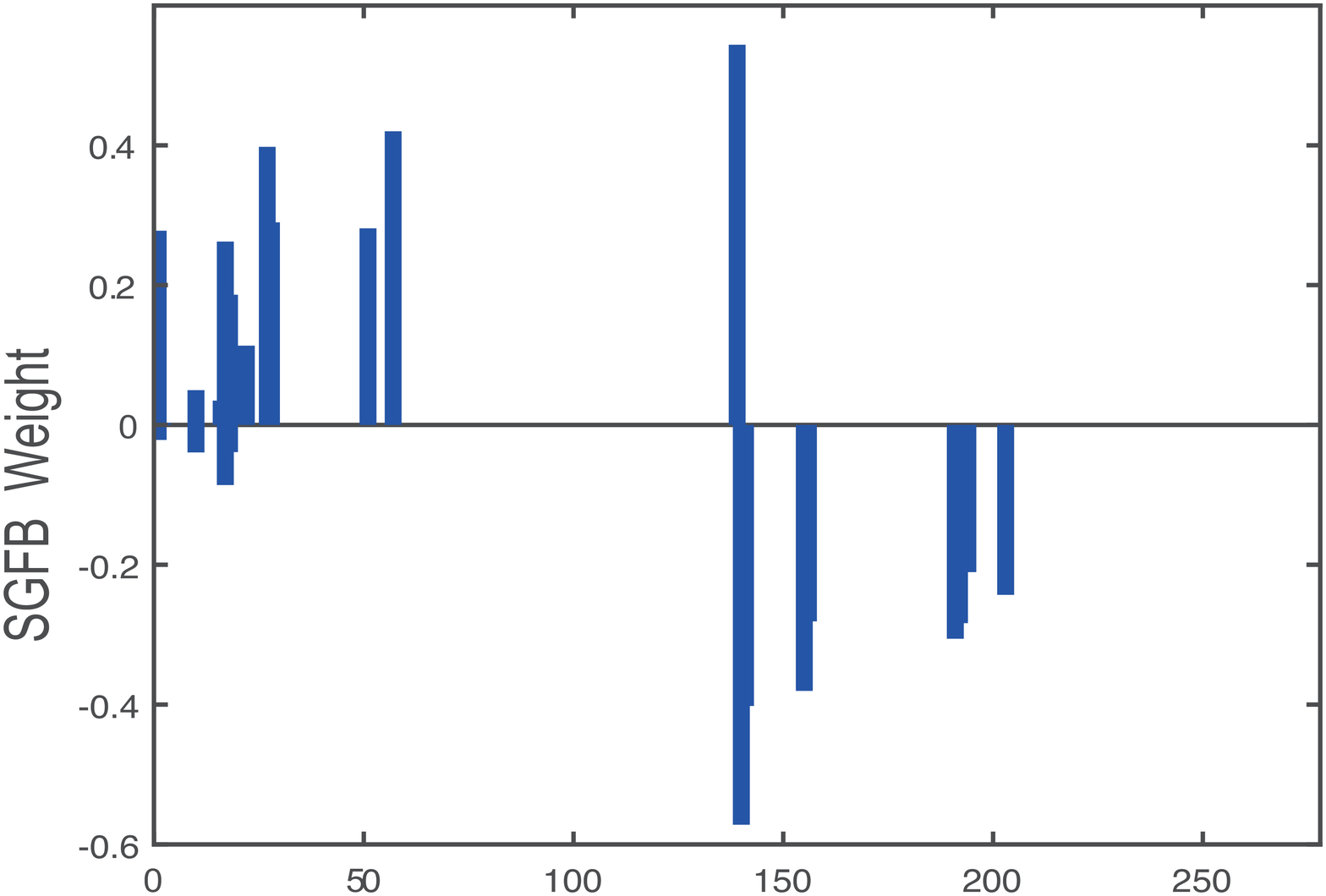}}
\subfigure[]{\includegraphics[scale=0.13]{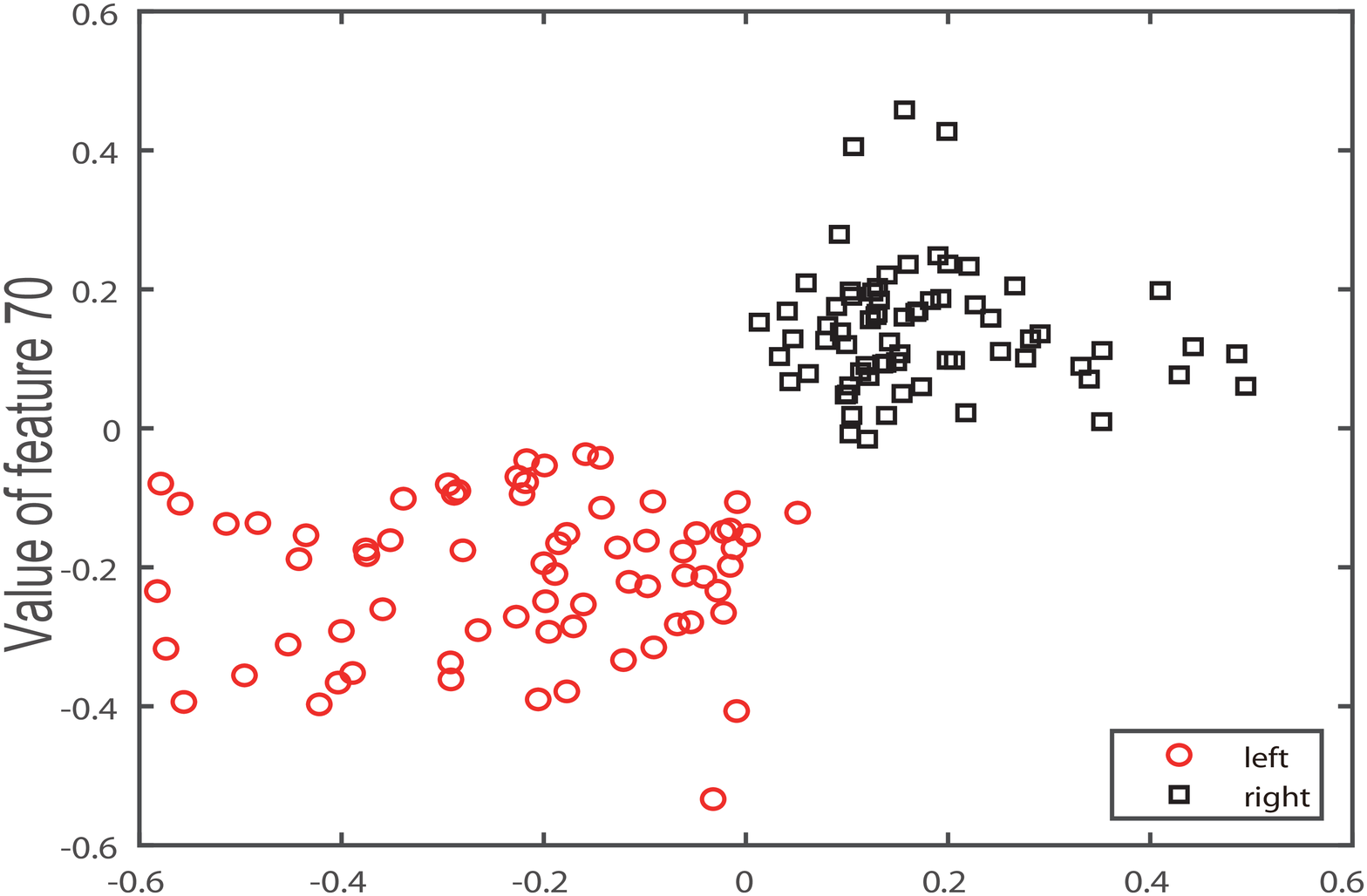}}
\subfigure[]{\includegraphics[scale=0.54]{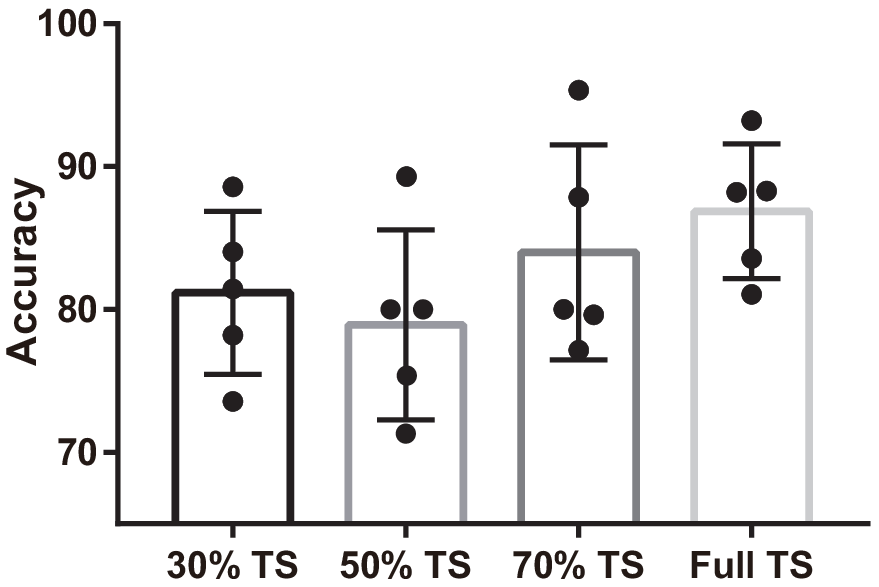}}
\caption{Performance of different classification tasks. (a)
Discriminant vectors $u^{*}$ and feature distributions of subject
ay, derived by SGFB. The vertical axis of the upper and lower
subgraphs represents the weight of the discrimination vector. Large
weights indicate that the corresponding features are more important
for MI classification. (b)The scatter plot of two category features.
(c) Comparison of the recognition accuracy for different TS
situations. \label{fig:3}}
\end{figure*}

\begin{table}[tp]
  \centering
  \fontsize{9.1}{9.1}\selectfont
  \caption{Recognition accuracy (\%) obtained by (C3-C4)+SVM, CSP+SVM (8-12Hz), CSP+SVM (4-40),
  FBCSP+SVM (multi-subband) and SGFB (multi-subband), respectively, at
 different time window (TW), for the BCI Competition IV dataset IVb.}
  \label{tab:performance_comparison}
\setlength{\tabcolsep}{2.15 mm}{
\begin{tabular}{lcllllll}\toprule
\multirow{2}{*}{~~~TW} &
\multicolumn{1}{l}{\multirow{2}{*}{~~~~~~~~~~~~~~~~~~Method}} &
\multicolumn{5}{c}{Subject} &
\multicolumn{1}{c}{\multirow{2}{*}{Average}} \\\cmidrule(lr){3-7}
                    & \multicolumn{1}{l}{}                                                                                           &~~aa       &~~al       &~~av       &~~aw       &~~ay       & \multicolumn{1}{c}{}\\\midrule
\multirow{6}{*}{1-2 (1s)}  & \begin{tabular}[c]{@{}c@{}}(C3-C4)+SVM (8-12 Hz)\end{tabular}            &54.62     &57.36     &57.82     &59.31     &54.47     &56.71                      \\
                           & \begin{tabular}[c]{@{}c@{}}CSP+SVM (8-12 Hz)\end{tabular}                          &75.52     &~79.1      &73.67     &76.33     &71.05     &75.13                      \\
                           & \begin{tabular}[c]{@{}c@{}}CSP+SVM (4-40 Hz)\end{tabular}                         &~79.9      &~84.4     &~89.2     &80.61     &~80.7      &84.96                      \\
                           & \begin{tabular}[c]{@{}c@{}}FBCSP+SVM (multi-subband)\end{tabular}             &55.91     &79.58     &63.25     &73.53     &78.56     &70.17                      \\
                           & \begin{tabular}[c]{@{}c@{}}SGFB (multi-subband)\end{tabular}                       &\bf83.57  &\bf95.36  &\bf88.21  &\bf81.07  &\bf93.21  &\bf88.28           \\\midrule
\multirow{6}{*}{2-3 (1s)}  & \begin{tabular}[c]{@{}c@{}}(C3-C4)+SVM (8-12 Hz)\end{tabular}           &57.65     &73.5      &68.66     &~70.1      &72.67     &68.52                      \\
                           & \begin{tabular}[c]{@{}c@{}}CSP+SVM (8-12 Hz)\end{tabular}                         &~66.9      &87.65    &71.85     &~82.5      &~~82        &77.98                      \\
                           & \begin{tabular}[c]{@{}c@{}}CSP+SVM (4-40 Hz)\end{tabular}                        &~77.7  &~78.2   &74.44     &77.12     &78.57     &~77.2                         \\
                           & \begin{tabular}[c]{@{}c@{}}FBCSP+SVM (multi-subband)\end{tabular}            &55.85     &61.19     &60.05     &65.24     &71.46     &62.78                         \\
                           & \begin{tabular}[c]{@{}c@{}}SGFB (multi-subband)\end{tabular}                      &74.64     &\bf91.07  &\bf77.86  &\bf84.64  &\bf92.86  &\bf84.21                \\\bottomrule
\end{tabular}}
\end{table}

\begin{table}[tp]
  \centering
  \fontsize{9.1}{9.1}\selectfont
  \caption{Classification results achieved by the SGFB for left and
right EEG signals (30\% TS).}
  \label{tab:4}
\setlength{\tabcolsep}{3.5 mm}{
\begin{tabular}{ccllllllll}\toprule
\multicolumn{1}{l}{\multirow{2}{*}{}}    &
\multicolumn{1}{l}{\multirow{2}{*}{evluation}} &
\multicolumn{5}{c}{Subject}   &
\multicolumn{1}{c}{\multirow{2}{*}{Average}} \\\cmidrule(lr){3-7}
\multicolumn{1}{l}{}                          &
\multicolumn{1}{l}{}& \multicolumn{1}{c}{aa}   &
\multicolumn{1}{c}{av}   &\multicolumn{1}{c}{aw}   &
\multicolumn{1}{c}{ay}   &\multicolumn{1}{c}{al}     &        &
\\\midrule
\multicolumn{1}{l}{\multirow{4}{*}} & \multicolumn{1}{l}{Acc}                         &~73.57         &~81.43         &~84.02           &~78.21       &~88.57          &~81.16         &                \\
\multicolumn{1}{l}{}                          & \multicolumn{1}{l}{Spec}              &~82.86         &~94.29         &~77.86           &~97.86       &~~~85           &~87.58         &                   \\
\multicolumn{1}{l}{}                          & \multicolumn{1}{l}{Sen}               &~64.29         &~68.57         &~90.71           &~58.57       &~92.14          &~74.86         &                    \\
\bottomrule
\end{tabular}}
\end{table}

\begin{table}[tp]
  \centering
  \fontsize{9.1}{9.1}\selectfont
  \caption{Classification results achieved by the SGFB for left and
right EEG signals (50\% TS).}
  \label{tab:5}
\setlength{\tabcolsep}{3.5 mm}{
\begin{tabular}{ccllllllll}\toprule
\multicolumn{1}{l}{\multirow{2}{*}{}}    &
\multicolumn{1}{l}{\multirow{2}{*}{evluation}} &
\multicolumn{5}{c}{Subject}   &
\multicolumn{1}{c}{\multirow{2}{*}{Average}} \\\cmidrule(lr){3-7}
\multicolumn{1}{l}{}                          &
\multicolumn{1}{l}{}& \multicolumn{1}{c}{aa}   &
\multicolumn{1}{c}{av}   &\multicolumn{1}{c}{aw}   &
\multicolumn{1}{c}{ay}   &\multicolumn{1}{c}{al}     &        &
\\\midrule
\multicolumn{1}{l}{\multirow{4}{*}} & \multicolumn{1}{l}{Acc}                         &~75.36         &~ 79.29        &~71.43         &~79.29       &~89.29         &~ 79.83        &                \\
\multicolumn{1}{l}{}                          & \multicolumn{1}{l}{Spec}              &~86.59         &~ 97.86        &~73.57         &~94.29       &~85.71         &~~  87.6       &                   \\
\multicolumn{1}{l}{}                          & \multicolumn{1}{l}{Sen}               &~~~70          &~ 60.71        &~69.29         &~64.29       &~92.86         &~  71.43       &                    \\
\bottomrule
\end{tabular}}
\end{table}

\begin{table}[tp]
  \centering
  \fontsize{9.1}{9.1}\selectfont
  \caption{Classification results achieved by the SGFB for left and
right EEG signals (70\% TS).}
  \label{tab:6}
\setlength{\tabcolsep}{3.5 mm}{
\begin{tabular}{ccllllllll}\toprule
\multicolumn{1}{l}{\multirow{2}{*}{}}    & \multicolumn{1}{l}{\multirow{2}{*}{evluation}} & \multicolumn{5}{c}{Subject}   & \multicolumn{1}{c}{\multirow{2}{*}{Average}} \\\cmidrule(lr){3-7}
\multicolumn{1}{l}{}                          & \multicolumn{1}{l}{}& \multicolumn{1}{c}{aa}   & \multicolumn{1}{c}{av}   &\multicolumn{1}{c}{aw}   & \multicolumn{1}{c}{ay}   &\multicolumn{1}{c}{al}     &        &  \\\midrule
\multicolumn{1}{l}{\multirow{4}{*}} & \multicolumn{1}{l}{~~Acc}                &~~79.64          &~~~80           &~~77.14          &~87.86     &~95.36         &~~~~84       &                 \\
\multicolumn{1}{l}{}                          & \multicolumn{1}{l}{~~Spec}           &~~69.14          &~~~90           &~~57.86          &~99.29     &~97.86         &~~82.83      &                 \\
\multicolumn{1}{l}{}                          & \multicolumn{1}{l}{~~Sen}            &~~92.14          &~~~65           &~~96.43          &~76.43     &~92.86         &~~84.57      &                 \\
\bottomrule
\end{tabular}}
\end{table}

\begin{table}[tp]
  \centering
  \fontsize{9.1}{9.1}\selectfont
  \caption{Classification results achieved by the SGFB for left and
right EEG signals (Full TS).}
  \label{tab:7}
\setlength{\tabcolsep}{3.5 mm}{
\begin{tabular}{ccllllllll}\toprule
\multicolumn{1}{l}{\multirow{2}{*}{}}   &
\multicolumn{1}{l}{\multirow{2}{*}{evluation}} &
\multicolumn{5}{c}{Subject} &
\multicolumn{1}{c}{\multirow{2}{*}{Average}} \\\cmidrule(lr){3-7}
\multicolumn{1}{l}{}                          &
\multicolumn{1}{l}{}& \multicolumn{1}{c}{aa}   &
\multicolumn{1}{c}{av}   &\multicolumn{1}{c}{aw}   &
\multicolumn{1}{c}{ay}   &\multicolumn{1}{c}{al}     &        &
\\\midrule
\multicolumn{1}{l}{\multirow{4}{*}} & \multicolumn{1}{l}{~~Acc}                   &~~83.57           &~~88.21          &~~81.07         &~93.21     &~95.36       &~~88.28        &                  \\
\multicolumn{1}{l}{}                          & \multicolumn{1}{l}{~~Spec}             &~~80.07           &~~82.66          &~~98.57         &~92.14     &~91.43       &~~88.97        &                  \\
\multicolumn{1}{l}{}                          & \multicolumn{1}{l}{~~Sen}              &~~86.43           &~~83.57          &~~63.57         &~94.29     &~99.29       &~~85.43        &                  \\
\bottomrule
\end{tabular}}
\end{table}

\begin{figure*}[t]\centering
\psfrag{t}[c][c][0.7]{$t$ (s)}
\psfrag{hc1}[c][c][0.65]{~~$\hat{c}_1$}
\psfrag{hc2}[c][c][0.65]{$\hat{c}_2$}
\subfigure[]{\includegraphics[scale=0.545]{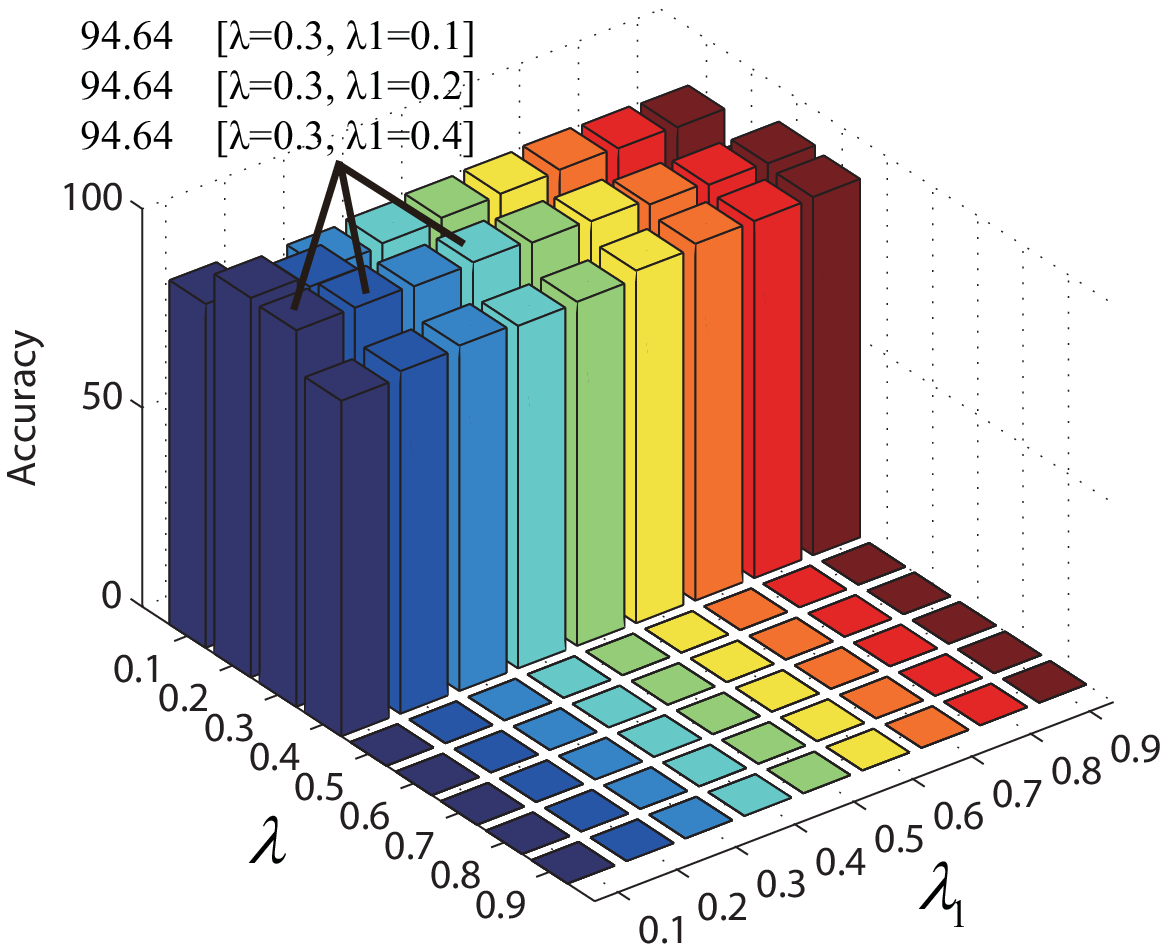}}
\subfigure[]{\includegraphics[scale=0.545]{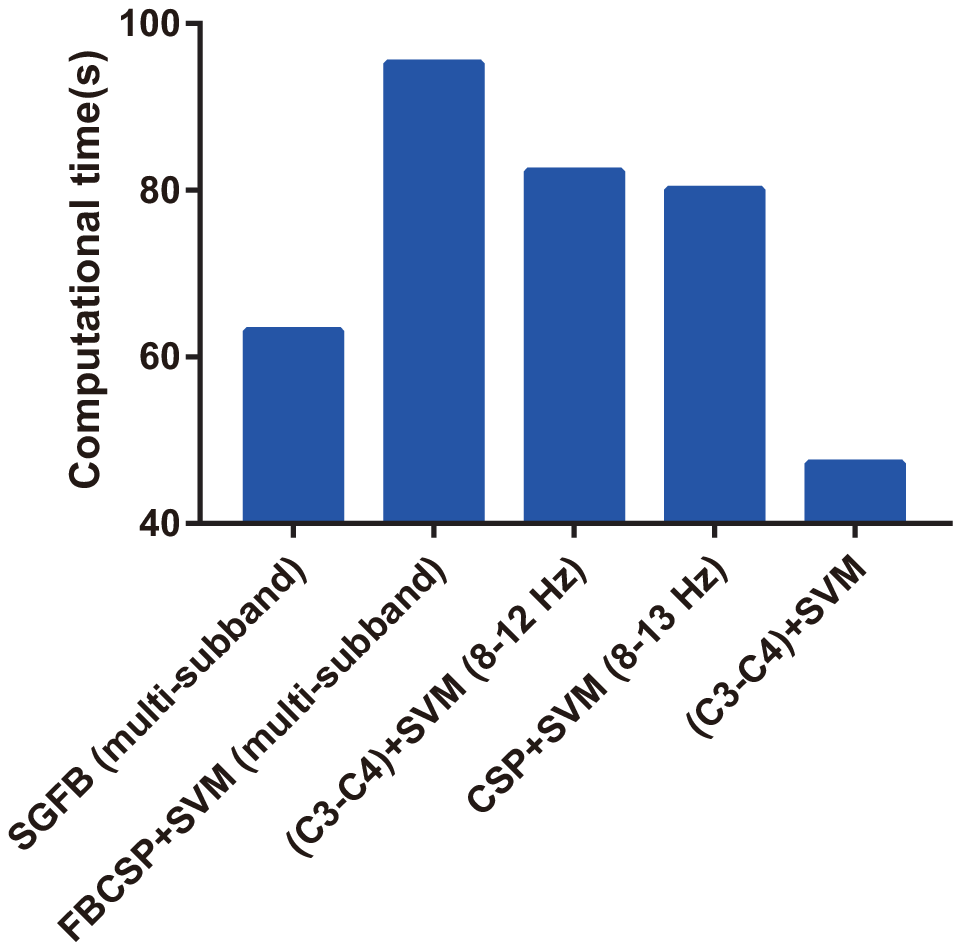}} \caption{(a).
Classification accuracy with different values of parameters
$\lambda$ and $\lambda_{1}$ respectively. The results indicate that
the performance of our method slightly fluctuates when the values of
hyperparameters vary in a relatively large range. In most cases, the
classification accuracies are generally stable with respect to the
changes of the two hyperparameters, indicating the robustness of our
proposed SGFB algorithm to the parameter values. (b). Average
computational time of different methods including SGFB, FBCSP+SVM
(multi-subband), (C3-C4)+SVM, CSP+SVM (8-13Hz) and (C3-C4)+SVM
respectively. \label{fig:4}}
\end{figure*}

\subsection{Parameter sensitivity}
\label{sec:parameter} In order to evaluate the effectiveness of the
proposed EEG classification method and superior to the existing
methods, five other different algorithms are compared to the
proposed algorithm using the same dataset.

The 10 Fold cross validation (10FCV) scheme is adopted for
evaluation of algorithm performance. In each fold of 10FCV
procedure, an additional inner 10FCV is also carried out on the
training data to determine the optimal hyperparameters (i.e.,
$\lambda$ for SFBG and the soft-margin parameter C for SVM). The
selection ranges of $\lambda$ and $\lambda_{1}$ are $[0.1, 0.2,
\cdots, 0.9]$ and $\text[0.1, \cdots, 0.4]$ respectively, while C is
selected from $[0.05, 0.1, \cdots, 1]$. The effectiveness of our
proposed method is affected by the selection of hyperparameters,
i.e., $\lambda$ for weighted group sparsity and $\lambda_{1}$ for
the pattern similarity. In our experiment, we implement a grid
search to select the optimal parameter values on the training data
using inner 10FCV. To investigate the parameter sensitivity of our
proposed algorithm, we evaluate effects of varying values of these
two hyperparameters on classification accuracy using 10FCV with all
subjects. The best accuracy of 94.64\% is achieved by using
$\lambda$ = 0.3 for strength weighted sparsity and $\lambda_{1}=
0.1$ for similarity constraint. Our future study will further
validate performance of the proposed algorithm on a completely
independent dataset.

\subsection{Computational efficiency}
We also compare the computational efficiency of above-mentioned
algorithms. Fig. \ref{fig:4}-b shows the calculation time evaluated
in a MATLAB R2019a environment on a desktop computer with a 3.8 Ghz
CPU (i7-7700k, 16G RAM). It suggests that these algorithms can be
run on internal loop cross-validation, which is needed for training
models. We note that the FBCSP spent the longest calculation time on
MI classification. The proposed SGFB algorithm shows comparable
computational efficiency with the best accuracy and comparable
computational efficiency. At present, in order to explore more
complex relationships in the feature space, many subspace
regularization algorithms have appeared for various applications
including imaging processing and encephalopathy diagnosis
\cite{Wu2015,Zhu2016aa,Zhou2016,Liu2018,Ahmadlou2012,Ahmadlou2012a,Ahmadlou2011,Ahmadlou2011a}.
A popular subspace learning method provides an effective method for
characterizing feature distributions and has recently been used to
improve the correlation of a single event latent classification
\cite{Higashi2016,Deng2018aa}. This kind of relationship constraint
based on manifold learning can further improve the accuracy of
classification, which is worthy of our future research
\cite{Zeng2018,zhu2014,Figueiredo2003}.

\section{Conclusion}
\label{sec:Con} In this paper, we introduce a novel SGFB algorithm
to construct an efficient classification model in motor
imagery-based BCI applications. We construct a compound matrix
dictionary for the frequency bands of different rhythms. By
combining $L_{2}$-norm and $L_{1}$-norm, a sparse group
representation model is designed to control the sparseness between
frequency bands to automatically determine the best model. With the
public motor imagery-related EEG datasets, extensive experimental
comparisons are carried out between the proposed algorithm and two
other state-of-the-art methods. The results illustrate the
effectiveness of our method in motor imagery BCI applications.

\section{Conflict of interest}
We declare that we do not have any commercial or associative interest that
represents a conflict of interest in connection with the work submitted.

\section{CRediT author statement}
{\bf Cancheng Li:} Conceptualization, Methodology, Software, Formal
analysis, Investigation, Writing, Original Draft, Visualization.
{\bf Chuanbo Qin:} Investigation, Visualization, Writing Review.
{\bf Cancheng Li:} Formal analysis, Investigation, Visualization,
Writing Review. {\bf Jing Fang:} Writing Review, Resources, Data
Curation, Supervision.

\section{Acknowledgments}
The authors would like to thank the editors and anonymous reviewers
for their valuable suggestions and constructive comments, which have
really helped the authors improve very much the presentation and
quality of this paper. The author would like to acknowledge and
thank the organizers of the BCI Competitions III.

\section{appendix}
\begin{algorithm}[!t]
  \label{algorithm:1} \caption{Pseudo-code Implementation of the
  Sparse Group Filter Bank Representation (SGFB) Model for Motor
  Imagery EEG Classification.}
  \begin{algorithmic}[1]
  \REQUIRE ~~ \\
  $\mathbf{\tilde D =\left[
  {{\tilde D_{f1}},{\tilde D_{f2}}, \cdots ,{\tilde
  D_{f9}}}\right]}$, hyperparameter $\lambda$ and
  $\lambda_{1}$.\\
  \ENSURE  ~~     \\
  The optimal sparse matrix ${u^*} = {u^{(L + 1)}}
  = \left[ {u_1^{(L + 1)},u_2^{(L + 1)}, \cdots ,u_n^{(L + 1)}}
  \right]$\\
  \STATE Initialize the sparse coefficients and its corresponding
  sign set;\\
  \FOR {$i \leftarrow 1, \cdots,  L$ }
  \FOR {$i \leftarrow 1, \cdots,  T$ }
        \IF{$the~coefficient~c_i^{(n)} = 0$}
        \STATE choose $j = \mathop {\arg \max }\limits_i \left| {\frac{{\partial {{\left\| {y - \widetilde D{u^*}} \right\|}^2}}}{{\partial u_j^*}}}    \right|$;\\
        \IF{$\frac{{\partial {{\left\| {y - \widetilde D{u^*}} \right\|}^2}}}{{\partial u_j^*}} >\lambda$}
        \STATE set $c_i^{j(n)} =  - 1$;\\
        \ELSIF{$\frac{{\partial {{\left\| {y - \widetilde D{u^*}} \right\|}^2}}}{{\partial u_j^*}} <\lambda$}
       \STATE set $c_i^{j(n)} =  1$;\\
       \ENDIF
        \ENDIF
  \ENDFOR\STATE {\bf Set feature sign search step} according to Algorithm 2\\
  \ENDFOR \RETURN coefficient $s$.
  \end{algorithmic}
  \end{algorithm}

  \begin{algorithm}[!t]
  \label{algorithm:3} \caption{Set feature sign search step.}
  \begin{algorithmic}[1]
  \STATE  (a)~~$\widehat {{u^*}}$ is a sub-column vector corresponding to the active set in the\\
  ~~~~~~complete atomic dictionary ${u^*}$\\
  \STATE  (b)~~Calculate the solution of Eq. \ref{equ:11}\\
  ~~~~~~~$\mathop {\arg\min}\limits_{{{\widehat u}_i}} \left\| {y - \widetilde D{u^*}}\right\| + \lambda \widehat \theta {\widehat u_i}$;\\
  ~~~~~~~$\widehat u_i^{\left( n \right)} = {\left( {{{\widehat u}^{{*^T}}}T{{\widehat u}^*}} \right)^{ - 1}}\left( {{{\widehat u}^{{*^T}}}y - \frac{{\lambda \widehat \theta }}{2}} \right)$;\\
  \STATE (c)~~Perform a discrete linear search on the close line segment\\
  ~~~~~~from $\widehat u$ to ${\widehat u_{new}}$\\
  \STATE~~~~~~ (1) Check the target function value of ${\widehat u_{new}}$ and the point in\\
  ~~~~~~~~~~~~~which all coefficient;\\
  \STATE~~~~~~ (2) Update the value of $\widehat u$ to the point with the smallest\\
  ~~~~~~~~~~~~~ objective function value.\\
  \STATE  (d)~~remove the coefficient of zero and update $\theta  = {\mathop{\rm sgn}} \left( {\widehat u_i^{*\left( n \right)}} \right)$\\
  ~~~~~~in the active set;\\
  \IF{$\forall~ u_j^* \ne 0$ (for non-zero coefficients)}
  \STATE $\frac{{\partial {{\left\| {y - \widetilde D{u^*}} \right\|}^2}}}{{\partial u_j^*}} + \lambda sign\left({u_j^*} \right)=0$;\\
  \ENDIF
  \IF{$\forall~ u_j^* = 0$ (for zero coefficients )}
  \STATE $\left| {\frac{{\partial \left\| {y -\widetilde D{u^*}} \right\|}}{{\partial u_j^*}}} \right| < \lambda$;\\
  \ELSE
  \STATE continue feature-sign search, go to step (b);
  \ENDIF
  \label{algorithm:3}
  \end{algorithmic}
  \end{algorithm}

  \begin{algorithm}[!t]
  \label{algorithm:2} \caption{SGFB for the EEG Classification.}
  \begin{algorithmic}[1]
  \REQUIRE ~~ \\
  coefficient s and the test sample signal $\mathbf{y} \in {\mathbb{R}^{m\times1}}$.\\
  \ENSURE  ~~     \\
  $class(\overline y) = \mathop {\arg \min }\limits_i {r_i}\left( {\overline y }\right)$.\\
  \STATE Normalize $\tilde D$ and $\overline y$ according to
  Eq.(\ref{equ:24})\\
  \STATE The optimal representation vector
  $\mathbf{u^{*}}$ according to Algorithm 1\\
  \STATE Compute ${\rm{class}}\left( \bar{y} \right) = \mathop {\arg
  \min }\limits_i {r_i}\left(\bar{y} \right)$\;
  \end{algorithmic}
  \end{algorithm}

\end{document}